\documentstyle{article}

\def\={\discretionary{-}{}{-}}

\def\ds{\displaystyle}

\def\astretch#1{\renewcommand{\arraystretch}{#1}}
\def\caka{Cartan-K\"ahler Theorem}

\def\cc{Cartan character}

\def\de{differential equation}

\def\hp{Hilbert polynomial}
\def\intc{integrability condition}

\def\ode{ordinary \de}
\def\pde{partial \de}

\def\tc{Taylor coefficient}
\def\aq#1#2{\alpha_{#1}^{(#2)}}

\def\d{{\rm d}}

\def\diag{\mbox{diag}\,}
\def\diff#1#2{\frac{\partial#1}{\partial#2}}

\def\Rc#1{{\cal R}_{#1}}

\newlength{\rulen}     
\newlength{\ruoff}     
\newlength{\rebox}     

\begin{document}

\title{Quantization of the Reduced Phase Space of
       Two-Dimensional Dilaton Gravity}
\author{Werner M. Seiler\thanks{On leave from Institut f\"ur Algorithmen
          und Kognitive Systeme, Universit\"at Karls\-ruhe, 76128
          Karlsruhe, Germany}, Robin W. Tucker\\
           School of Physics and Materials\\
           Lancaster University\\
           Bailrigg, LA1 4YB, UK\\
           Email: W.Seiler@lancaster.ac.uk, R.W.Tucker@lancaster.ac.uk}
\date{}
\maketitle

\begin{center}
\bf Abstract
\end{center}
\begin{quote}

{\small We study some two-dimensional dilaton gravity models using the formal
theory of \pde s. This allows us to prove that the reduced phase space is
two-dimensional without an explicit construction. By using a convenient
(static)
gauge we reduce the theory to coupled \ode s and we are able to derive for some
potentials of interest closed-form solutions. We use an effective (particle)
Lagrangian for the reduced field equations in order to quantize the system in a
finite-dimensional setting leading to an exact partial differential
Wheeler-DeWitt equation instead of a functional one. A {\small WKB}
approximation for some quantum states is computed and compared with the
classical Hamilton-Jacobi theory. The effect of minimally coupled matter is
examined.}
\end{quote}

\section{Introduction}

A tensor formulation of physical theories makes no overt reference to any
particular frame of reference. To interpret these theories it is often
necessary
to extract from them coordinate independent information. In particular in
theories of classical gravitation, coordinate freedom is either explicitly
removed by working in a particular coordinate system or regarded as generating
constraints for the subsequent analysis.  For the classical theory it may
simply
be a matter of taste as to which procedure is adopted. However the discussion
of
quantization is often acutely sensitive to the choice adopted.

Following Witten's observation~\cite{wit:black} that models of two dimensional
dilaton gravity offer a means of studying the Hawking effect with back reaction
there has been an enormous interest in such models. They arise naturally from
certain truncations of low energy string effective actions~\cite{cghs:black}
and
symmetric configurations in higher dimensions~\cite{tk:ssg}.  Such models have
been rendered completely integrable at the classical level by exploiting the
local conformal flatness of all two dimensional manifolds and their
quantization
discussed from several alternative viewpoints.

In this paper we reexamine the conditions that are responsible for this
remarkable integrability and offer an alternative quantization. The basic
observation is that a particular conformal gauge reduces the classical
integrability to the problem of solving a system of {\em ordinary\/} \de s.
The
existence of this gauge is proved using methods from the formal theory of \pde
s~\cite{pom:eins,wms:diss}.  This approach allows us to compute the dimension
of
the reduced phase space without explicitly constructing it.  This technique
should also prove useful in more complicated theories where explicit reduction
is not possible.

Using methods from the Hamilton-Jacobi theory for systems with constraints
\cite{ht:quant,dlgp:hjc} we construct local expressions for the dynamical
degrees of freedom for dilaton gravity on the line.  This is in marked
difference to other approaches to the quantization of the reduced
theory~\cite{lmgk:dilaton,nsnta:grav2d}. The quantum amplitudes are shown to
satisfy a simple hyperbolic wave equation which is exactly soluble for
appropriate boundary conditions.  Similar quantum theories were already
obtained
by different authors~\cite{nsna:wave,ot:cqd} in an approximate minisuperspace
approach.  But here it is not necessary to make such an approximation because
of
the finite-dimensional reduced phase space.

A straightforward semi-classical analysis of the exact quantum description
yields a {\small WKB} phase that encodes all the classical dilaton gravity
solutions. We explicitly demonstrate that the integral curves that annihilate
the gradient of the {\small WKB} phase form a family of exact classical vacuum
solutions. This suggests that such a quantization of dilaton gravity deserves
further scrutiny.

The article is organized as follows: After a brief discussion of the classical
action and its field equations, we use in Section~\ref{formana} a formal
analysis to derive indirectly the dimension of the reduced phase space. In
Section~\ref{redode} we explicitly reduce the field equations by a gauge fixing
to a system of \ode s and construct its general solution. After considering
some
explicitly solvable models we proceed in Section~\ref{hjt} to the
Hamilton-Jacobi analysis of the system. Its results are used in
Section~\ref{quant} for the quantization. Section~\ref{matter} discusses the
effect of minimally coupling a matter field. Finally, some conclusions are
given.

\section{Classical Action and Field Equations}\label{action}

In two dimensions a general coordinate invariant Lagrangian density containing
the metric~$g_{\mu\nu}$, a scalar dilaton field~$\Phi$ and their derivatives up
to second order is given by
\begin{equation}\label{dggenact}
{\cal L}[g_{\mu\nu},\Phi]=
\sqrt{-g}\left(\frac{1}{2}g^{\mu\nu}\partial_\mu\Phi\partial_\nu\Phi+
V(\Phi)+D(\Phi)R\right)
\end{equation}
where $R$ denotes the curvature scalar associated with the metric and $D$ is a
scalar function.

Using field redefinitions one can, however, considerably simplify this action.
The kinetic term for~$\Phi$ can be eliminated by a Weyl rescaling of the
metric~\cite{lmgk:dilaton}
\begin{equation}
\bar g_{\mu\nu}=\Omega^2(\Phi)g_{\mu\nu}\,,
\end{equation}
if $\Omega$ satisfies the differential equation
\begin{equation}\label{omega}
4D^\prime(\Phi)\frac{\d\ln\Omega}{\d\Phi}=1\,.
\end{equation}
If we additionally redefine the dilaton field $\bar\Phi=D(\Phi)$, we obtain the
action
\begin{equation}\label{dgact}
{\cal L}[\bar g_{\mu\nu},\bar\Phi]=\sqrt{-\bar g}\,\Bigl(\bar
V(\bar\Phi)+\bar\Phi \bar R\Bigr)
\end{equation}
where the new potential $\bar V(\bar\Phi)$ is given by
\begin{equation}\label{barV}
\bar V(\bar\Phi)=\frac{V(\Phi(\bar\Phi))}{\Omega^2(\Phi(\bar\Phi))}\,.
\end{equation}
(One must be careful here, if $D$ has critical points~\cite{bl:dilaton}).

Henceforth we will restrict our attention to this action and drop the bar over
the fields.  Variation with respect to the metric yields the ``Einstein
Equations''
\begin{equation}\label{eineqn}
\nabla_\mu\nabla_\nu\Phi-
g_{\mu\nu}\Bigl[\nabla^\rho\nabla_\rho\Phi+V(\Phi)\Bigr]=0\,,
\end{equation}
whereas variation with respect to~$\Phi$ leads to the additional equation
\begin{equation}\label{phieqn}
R+V^\prime(\Phi)=0
\end{equation}
determining the curvature scalar.

Before we start a detailed analysis of these field equations, we study briefly
the relation between the potentials appearing in (\ref{dggenact})
and~(\ref{dgact}) for the most often considered case $D(\Phi)=a\Phi^n$ for some
constants~$a,n$. If $n\neq2$ a solution of~(\ref{omega}) is given by
\begin{equation}
\Omega(\Phi)=e^{-\Phi^{2-n}/4an(n-2)}\,;
\end{equation}
while for $n=2$
\begin{equation}
\Omega(\Phi)=\Phi^{1/8a}\,.
\end{equation}

For $n=1$ we obtain thus from~(\ref{barV})
\begin{equation}\label{n1}
\bar V(\bar\Phi)=V(\bar\Phi/a)e^{-\bar\Phi/2a^2}\,.
\end{equation}
This implies especially that for an exponential potential $V(\Phi)\sim
e^{\nu\Phi}$ the potential remains an exponential after the transformation but
with a modified coefficient $\bar\nu=(2a\nu-1)/2a^2$. Note that this results
also holds for $\nu=0$, i.e.\ if the potential consists just of a cosmological
constant.  Conversely, the potential becomes constant, if $a=1/2\nu$.

For $n=2$ the transformation reads
\begin{equation}
\bar V(\bar\Phi)=V(\pm\sqrt{\bar\Phi/a})(\bar\Phi/a)^{-1/8a}\,.
\end{equation}
Thus Lagrangian densities of the form $\bar {\cal L}=\bar\Phi
R+\Lambda\bar\Phi^n$ as they are e.g.\ considered in Ref.~\cite{mig:dilaton}
can
be derived from a model in the form~(\ref{dggenact}) with $D(\Phi)=-\Phi^2/8n$
and a ``cosmological constant'' as potential $V(\Phi)=\Lambda/(-8n)^n$.

A class of models which appeared first in effective string actions and which
has
found considerable interest due to the existence of black hole
solutions~\cite{cghs:black,wit:black} is described by the action
\begin{equation}\label{cghs}
{\cal L}[g_{\mu\nu},\varphi]=\frac{1}{8}\sqrt{-g}e^{-2\varphi}
\left(R+4(\nabla\varphi)^2+c\right)
\end{equation}
where $c$ is a constant.  Using field redefinitions one can transform it
to~\cite{rt:dilaton}
\begin{equation}
\bar{\cal L}[\bar g_{\mu\nu},\phi]=\sqrt{-\bar g}\left(
\frac{1}{2}\bar g^{\mu\nu}\partial_\mu\phi\partial_\nu\phi+
\frac{1}{2}q\phi\bar R+\frac{1}{8}c e^{\phi/q}\right)
\end{equation}
with an arbitrary constant~$q$.  Elimination of the kinetic term leads then to
a
modification of the exponential.  Note that this simple Liouville form of the
transformed action is due to the factor~4 in~(\ref{cghs}). A different
factor~$\gamma$ leads to a modified potential of the form
$\mu\phi^{1-\gamma/4}e^{\phi/q}$.

\section{Formal Analysis}\label{formana}

The first step in a formal analysis is always to complete the given system of
\pde s to an involutive one~\cite{pom:eins,wms:diss}. This completion is
closely
related to the Dirac formalism for systems with constraints. Actually, one can
interpret the Dirac algorithm as a completion procedure for the Hamilton-Dirac
equations of the system~\cite{wms:con1}.

In our case the involution analysis is rather simple, as it is straightforward
to show that the combined field equations~(\ref{eineqn},\ref{phieqn}) are
already in involution. An interesting fact hereby is that (\ref{eineqn})
entails~(\ref{phieqn}), if we exclude the trivial case that $\Phi$ is constant.
The \intc s of~(\ref{eineqn}) require that either (\ref{phieqn})~holds or
$\Phi$~must be constant. Similar effects are known from other theories coupled
to gravity.

The arbitrariness of the general solution of a system of $q$-th order \pde s in
$n$~independent variables can be determined from its \cc s~$\aq{q}{k},\
k=1,\dots,n$~\cite{wms:str}. A simple calculation for our system yields
\begin{equation}\label{cc}
\aq{2}{2}=2\,,\quad\aq{2}{1}=6\,.
\end{equation}
By a comparison with a Taylor expansion of the general solution these
characters
can be interpreted in terms of numbers of arbitrary functions of different
numbers of arguments. Here we obtain that the general solution of our field
equations can be written as an algebraic expression containing two arbitrary
functions of two arguments and two arbitrary functions of one argument.

Another way to represent the arbitrariness of the general solution is given by
the \hp~$H(r)$ of the field equations. It denotes the number of \tc s of
order~$r$ which can be chosen arbitrarily. From (\ref{cc}) we
obtain~\cite{wms:diss,wms:str} (note the slightly different notation used
there)
\begin{equation}
H(r)=2r+4\,.
\end{equation}
It is important to note that $H(r)$ yields the correct values only for
$r\geq2$,
as we are dealing with second-order equations. On the other hand the number of
arbitrary \tc s of order less than or equal to~2 is determined by the dimension
of the submanifold described by the field equations in the appropriate
second-order jet bundle; thus in our case it is~20.

We must, however, adjust for the covariance under coordinate transformations.
Especially the two functions of two variables stem obviously from this gauge
covariance. We have recently shown how such a correction can be performed as
soon as the gauge group is known~\cite{wms:sym,wms:str}. The key is the
introduction of a gauge corrected \hp\ which in turn leads to gauge corrected
\cc s.

In our case we must subtract the invariance under the transformation
\begin{equation}
\astretch{1.8}
\begin{array}{rcl}
g_{\mu\nu}&=&\ds
  \diff{\bar x^\rho}{x^\mu}\diff{\bar x^\sigma}{x^\nu}\bar g_{\rho\sigma}\,,\\
\Phi&=&\bar\Phi\,.
\end{array}
\end{equation}
The transformation depends on two gauge functions~$\bar x^\rho$ through their
first derivatives. Thus if we expand again in a power series, we can give
$G(r)$~coefficients of order~$r$ arbitrary values through gauge transformation
where $G(r)$ is given by
\begin{equation}
G(r)=2{r+2\choose r+1}=2r+4\,.
\end{equation}

By comparison with the \hp\ we see that all the arbitrariness for $r\geq2$
stems
from this gauge covariance. Hence the gauge corrected \cc s vanish and the
reduced phase space of this theory is finite-dimensional. Usually $G(r)$ yields
the correct values only from a certain value of~$r$ on. In our case, however,
one can easily see by writing out the first terms of the expansion that it is
correct for all~$r\geq0$. Thus we can further conclude that 18~\tc s of order
up
to two can be given arbitrary values by gauge transformation. Since the general
solution of our field equations contains only 20~arbitrary coefficients at
these
orders we obtain that the dimension of the reduced solution space is two. This
fact was also proven in Refs.~\cite{bl:dilaton,lmgk:dilaton} using an explicit
reduction.

Actually, in this simple case it is not necessary to use the \cc s to prove the
finiteness of the reduced phase space. The easier concept of the strength of a
\de\ introduced by Einstein~\cite{ein:rel,sue:str} suffices here. A
straightforward computation shows that the field equations are absolutely
compatible and have a vanishing strength, if one takes the gauge symmetry into
account. But since we are dealing with a two-dimensional space-time, this
implies immediately that the gauge reduced solution space is
finite-dimensional.
However, the exact dimension can be computed only with the refined analysis
used
above.

We can understand this finiteness by considering the metric as an external
field. (\ref{eineqn})~represents then a finite type system for the dilaton
field~$\Phi$, as each second order derivative of $\Phi$ is determined by an
equation. Thus the general solution of this system depends only on a finite
number of parameters. All arbitrary functions stem therefore from the metric as
solution of~(\ref{phieqn}).

\section{Reduction to Ordinary Differential Equations}\label{redode}

The solution of every system of finite type can be obtained by solving systems
of \ode s~\cite{cara:pdgl,lie:tg1}. The reduction is based on the theory of
complete systems and can be performed in a purely algorithmic way. However, in
our case it will not be necessary to follow this procedure which would lead to
a
fairly complicated system of \ode s~\cite{wms:diss}. By choosing an appropriate
gauge the reduction can be obtained directly.

We first exploit the well-known fact that every two-dimensional metric is
(locally) conformally flat~\cite{eis:rg} and set
\begin{equation}\label{conmet}
g_{\mu\nu}=e^{\lambda(x,t)}\eta_{\mu\nu}
\end{equation}
where $\eta_{\mu\nu}=\diag(-1,1)$ is the Minkowski metric.  The curvature
scalar
of such a metric is given by
\begin{equation}\label{R}
R=(\lambda_{tt}-\lambda_{xx})e^{-\lambda}\,.
\end{equation}
Thus after some trivial manipulations the combined field equations can be
written in the following form:
\begin{equation}\label{dgfeq}
\astretch{1.5}
\Rc{2}\,:\,\left\{\begin{array}{l}
\Phi_{tt}-\frac{1}{2}(\Phi_t\lambda_t+\Phi_x\lambda_x)+
\frac{1}{2}e^\lambda V(\Phi)=0\,,\\
\Phi_{xx}-\frac{1}{2}(\Phi_t\lambda_t+\Phi_x\lambda_x)-
\frac{1}{2}e^\lambda V(\Phi)=0\,,\\
\Phi_{xt}-\frac{1}{2}(\Phi_x\lambda_t+\Phi_t\lambda_x)=0\,,\\
\lambda_{tt}-\lambda_{xx}+e^\lambda V^\prime(\Phi)=0\,.
\end{array}\right.
\end{equation}

We show now that it suffices to consider static metrics. To prove that this
gauge is really attainable we apply the pseudogroup approach to Lie groups
based
on \de s~\cite{pom:eins}.  Assume we are given the metric
\begin{equation}
\bar g=e^{\omega(\xi,\tau)}(d\xi\otimes d\xi-d\tau\otimes d\tau)\,.
\end{equation}
By performing a change of coordinates $\xi=\xi(x,t)$, $\tau=\tau(x,t)$ we want
to obtain the metric~(\ref{conmet}) with $\lambda_t=0$. A straightforward
calculation shows that the following conditions
\begin{equation}\label{confgr}
\xi_x\xi_t=\tau_x\tau_t\,,\quad \xi_x^2-\tau_x^2=\tau_t^2-\xi_t^2
\end{equation}
are necessary and sufficient to maintain the conformal form. They form an
involutive system, as one can easily check.

The condition that the conformal factor is independent of~$t$ leads to the
additional equation
\begin{equation}\label{tcond}
2(\xi_t\xi_{tt}-\tau_t\tau_{tt})+
(\xi_t^2-\tau_t^2)(\omega_\xi\xi_t+\omega_\tau\tau_t)=0\,.
\end{equation}
Hence we must prove that the combined system~(\ref{confgr},\ref{tcond}) has a
solution for any function~$\omega$.

By a simple analysis of the pivots one can show that the symbol of the
second-order system obtained after prolonging~(\ref{confgr}) is involutive, if
$\xi_x\tau_t\neq \xi_t\tau_x$, and that furthermore no \intc s arise. But this
is exactly the condition for $\xi,\tau$ to define an invertible change of
coordinates. Hence the system is involutive and by the
\caka~\cite{bcggg:eds,pom:eins} we can always find a local solution~$\xi,\tau$
and construct from it a coordinate system with the required
property.\footnote{Using exactly the same technique one can also show that the
conformal form~(\ref{conmet}) can always be reached by a change of
coordinates. But one should perhaps remark that this proof is somewhat weaker
than the usual one found in textbooks on differential geometry, as the \caka\
holds only for analytic systems of \de s and guarantees only the existence of
analytic solutions.}

Prolonging the fourth equation of (\ref{dgfeq}) with respect to~$t$ under this
gauge condition yields that either $\Phi_t=0$ or the potential is linear
$V(\Phi)=A_1\Phi+A_2$. In the latter case the fourth equation of~(\ref{dgfeq})
obviously decouples from the system and we cannot obtain any information
about~$\Phi$ from our gauge condition. One can, however, show that the field
equations imply the existence of a Killing vector
\begin{equation}\label{kilv}
k_\mu=\epsilon_{\mu\nu}\nabla^\nu\Phi
\end{equation}
orthogonal to the gradient of~$\Phi$. Thus we can always choose the gauge
$\Phi_t=0$ which in turn leads to $\lambda_t=0$.

Thus we will assume from now on that we are in a coordinate system where
$\lambda_t=\Phi_t=0$.  The first two equations of~(\ref{dgfeq}) yield
$\Phi_{xx}-\Phi_x\lambda_x=0$.  Discarding the uninteresting case $\Phi_x=0$,
this can be integrated once and yields
\begin{equation}\label{phix}
\Phi_x=A e^\lambda
\end{equation}
with an integration constant~$A$. Note that this implies that the sign of
$\Phi_x$ never changes and that it is fixed by the sign of~$A$. Substituting
this in the second equation of~(\ref{dgfeq}) leads to
\begin{equation}\label{lambdax}
A\lambda_x=V(\Phi)\,.
\end{equation}
Differentiating (\ref{phix}) allows one to eliminate~$\lambda$ and arrive
finally at the simple equation
\begin{equation}\label{phixx}
A\Phi_{xx}-V(\Phi)\Phi_x=0\,.
\end{equation}
There is no need to consider the fourth equation of~(\ref{dgfeq}), as it is an
\intc\ and thus automatically satisfied.

Rewriting the potential as a derivative, $V(\Phi)=W^\prime(\Phi)$, one can
easily obtain an implicit solution of~(\ref{phixx}). Integrating once yields
the
first integral
\begin{equation}\label{fint}
A\Phi_x-W(\Phi)=B
\end{equation}
for some constant~$B$. Separation of variables leads to
\begin{equation}\label{impsol}
x(\Phi)+C=\int_0^\Phi\frac{A\,d\varphi}{B+W(\varphi)}\,.
\end{equation}
Once this expression is inverted to obtain $\Phi$ in explicit form,
$\lambda$~can be derived algebraically from~(\ref{phix})
\begin{equation}\label{lambphi}
\lambda=\ln\left(\frac{B+W(\Phi)}{A^2}\right)\,.
\end{equation}

We have thus found a three-dimensional solution space. This is no surprise, as
the field equations together with the used gauge conditions describe a
three-dimensional manifold in the second-order jet bundle. A similar
construction in light cone coordinates was presented in~\cite{lmk:birkhoff}.

To conclude this section we briefly discuss the three occurring integration
constants. $C$~can obviously be set to any value by changing the origin of the
coordinate system. Thus we can set it to zero without loss of generality.
Similarly, $A$~can be adjusted to any value by a coordinate scaling
$x\rightarrow x/A$, $t\rightarrow t/A$, as under such a transformation
$\lambda\rightarrow\lambda+\ln A^2$.

By contrast $B$ has an invariant meaning. Since $A=\Phi_x e^{-\lambda}$, we
obtain from~(\ref{fint})
\begin{equation}
B=e^{-\lambda}\Phi_x^2-W(\Phi)\,.
\end{equation}
This expression can be expressed covariantly as
\begin{equation}
B=g^{\mu\nu}\nabla_\mu\Phi\nabla_\nu\Phi-W(\Phi)\,.
\end{equation}
One can show that $B$ corresponds to the {\small ADM} energy of the
system~\cite{gk:qt}.

Thus only one of the three integration constants parameterizing the general
solution of the field equations has an invariant meaning. The other two can be
absorbed in coordinate transformation. This effect is extensively discussed in
Ref.~\cite{lmgk:dilaton}.

This is exactly the result one would expect in ordinary gravity from the
Birkhoff theorem: Up to coordinate transformations the static vacuum solutions
form a one-parameter family. For this reason some authors speak of the
generalized Birkhoff theorem of dilaton gravity~\cite{lmk:birkhoff}.

\section{Some Solvable Models}

We start by considering a linear potential of the form $V(\Phi)=k\Phi+m$, i.e.\
the so-called Jackiw-Teitelboim or Liouville gravity~\cite{jac:liou,tei:ham2d}
with a cosmological constant~$k\neq0$. In this case the field equations
decouple
and we obtain for the conformal factor the equation
\begin{equation}\label{pb}
\lambda_{xx}-k e^\lambda=0
\end{equation}
which can be considered either as a special case of the Poisson-Boltzmann
equation or as describing stationary solutions of the Liouville equation. Its
general solution is given by~\cite{jac:liou}
\begin{equation}
\lambda(x)=-\ln\left\{\frac{k}{2\beta^2}
                     \sinh^2\left[\beta(x-x_0)\right]\right\}
\end{equation}
with two integration constants~$\beta,x_0$. Obviously, $x_0$ is without
physical
significance and can be set zero. The curvature is constant
\begin{equation}
R=k\,.
\end{equation}
{}From (\ref{lambdax}) we obtain immediately
\begin{equation}
\Phi(x)=-\frac{A}{k}\left[2\beta\coth\left(\beta(x-x_0)\right)-\mu\right]\,.
\end{equation}

Next we consider potentials of the form $V(\Phi)=\alpha e^{\beta\Phi}$ as they
occur in the effective string actions.\footnote{As already mentioned in
Section~\ref{action} more generally one obtains a potential of the
form~$\alpha\Phi^\gamma e^{\beta\Phi}$. These models can still be solved
exactly~\cite{ls:black}; however, many case distinctions arise.} Here it is
simpler to go back to the equations~(\ref{phix},\ref{lambdax}) and to introduce
new dependent variables~$\psi,\mu$ by $\psi=V(\Phi)$ and $\mu=Ae^\lambda$. This
transformation yields the system
\begin{equation}
\psi_x=\beta\psi\mu=\beta A\mu_x\,.
\end{equation}
Thus these new variables are related through
\begin{equation}
\psi(x)=\beta A[\mu(x)+D]
\end{equation}
with an integration constant~$D$. Eliminating $\psi$ leads to a simple
Bernoulli equation for~$\mu$ which can be solved by separation of variables. We
must distinguish two cases: If $D=0$, we obtain
\begin{equation}
\mu(x)=\frac{1}{C-\beta x}
\end{equation}
and for the curvature scalar
\begin{equation}\label{RD0}
R=\frac{A\beta^2}{\beta x-C}
\end{equation}
with a further integration constant~$C$. Otherwise we find
\begin{equation}
\mu(x)=\frac{D}{Ce^{-\beta Dx}-1}
\end{equation}
and the curvature scalar
\begin{equation}\label{RDarb}
R=\frac{ADC\beta^2}{e^{\beta Dx}-C}\,.
\end{equation}
By setting $C=0$ in~(\ref{RD0}) and $C=1$ in~(\ref{RDarb}), respectively, we
can
move the singularity of the curvature to~$x=0$.

The third important model is provided by spherically symmetric gravity in
3+1~dimensions~\cite{tk:ssg}. It can be reduced to a dilaton gravity action in
two dimensions of the form~(\ref{dgact}) where the potential is given by
$V(\Phi)=1/\sqrt{2\Phi}$. As above we must distinguish two cases in the
integral
in~(\ref{impsol}). If $B=0$, the solution can be given in explicit form
\begin{equation}
\Phi(x)=\frac{1}{A^2}(x+C)^2\,.
\end{equation}
Otherwise an inversion is not possible. The implicit solution is
\begin{equation}
x+C=A\left[\sqrt{2\Phi}-B\ln\left(1+\sqrt{2\Phi/B^2}\right)\right]\,.
\end{equation}
In any case the curvature scalar is given by
\begin{equation}
R=\frac{1}{4}\Phi^{-3/2}\,.
\end{equation}

\section{Hamilton-Jacobi Theory}\label{hjt}

After the gauge reduction we obtained in Section~\ref{redode} the following
system of one first-order and two second-order \ode s
\begin{equation}\label{eqmg}
\astretch{1.3}
\begin{array}{l}
\Phi_{xx}-e^\lambda V(\Phi)=0\,,\\
\lambda_{xx}-e^\lambda V^\prime(\Phi)=0\,,\\
\Phi_x\lambda_x-e^\lambda V(\Phi)=0\,.
\end{array}
\end{equation}
Note that the first-order equation produces together with any of the
second-order ones the other second-order equation as an \intc. The two
second-order equations, however, form a normal system and thus cannot generate
the first-order one.

We now try to find an effective Lagrangian for the gauged equation of
motion~(\ref{eqmg}). A reasonable starting point is obtained by applying
our gauge conditions to the full Lagrangian density~(\ref{dgact}) and
integrating once by parts
\begin{equation}\label{lagg}
{\cal L}_g[\Phi,\lambda]=\Phi_x\lambda_x+e^\lambda V(\Phi)\,.
\end{equation}
The corresponding Euler-Lagrange equations are the two second-order equations
in~(\ref{eqmg}). Thus this action yields a too general dynamics, as it
``looses'' one condition! Performing a Legendre transformation on~(\ref{lagg})
shows that the missing equation demands the vanishing of the Hamiltonian of the
system (``zero energy condition'')
\begin{equation}\label{hamg}
{\cal H}_g=\pi_\Phi\pi_\lambda-e^\lambda V(\Phi)=0
\end{equation}
where the canonically conjugate momenta are given by $\pi_\Phi=\lambda_x$ and
$\pi_\lambda=\Phi_x$, respectively.

If we denote Hamilton's principal function as usual by $S$, the Hamilton-Jacobi
equation of the {\em unconstrained\/} system described by the Lagrangian~${\cal
L}_g$ is
\begin{equation}\label{hje}
\ds\diff{S}{x}+\diff{S}{\Phi}\diff{S}{\lambda}-e^\lambda V(\Phi)=0\,.
\end{equation}
Imposing the constraint~(\ref{hamg}) leads to a second equation for~$S$,
namely~\cite{ht:quant,dlgp:hjc}
\begin{equation}\label{hjc}
\diff{S}{\Phi}\diff{S}{\lambda}-e^\lambda V(\Phi)=0\,.
\end{equation}
Obviously, we can now discard~(\ref{hje}) by simply looking for a principal
function independent of~$x$.

Ideally, one would like to find a complete integral~$S(x,\Phi,\lambda,p_1,p_2)$
of~(\ref{hje}) such that it satisfies the constraint~(\ref{hjc}) for $p_2=0$.
Such a complete integral generates a canonical transformation to new
coordinates~$(q_1,q_2,p_1,p_2)$ via
\begin{equation}\label{ctrafo}
\astretch{2}
\begin{array}{ccc}
\ds\pi_\Phi=\diff{S}{\Phi}\,,&\quad&\ds\pi_\lambda=\diff{S}{\lambda}\,,\\
\ds q^1=\diff{S}{p_1}\,,&&\ds q^2=\diff{S}{p_2}\,.
\end{array}
\end{equation}
In these coordinates the system decouples~\cite{gnr:hjc} into an unconstrained
one depending only on the canonical pair~$(q^1,p_1)$ plus a trivial one
containing the gauge degree of freedom~$(q^2,p_2)$. $p_2$~is constrained to
zero
and $q^2$ remains completely arbitrary.

Unfortunately, we have not been able to construct such a complete integral.
However, we found an incomplete integral~\cite{ht:quant} satisfying the full
system~(\ref{hje},\ref{hjc})
\begin{equation}\label{inint}
S^{(0)}(\Phi,\lambda,p_1)=p_1 e^\lambda+\frac{W(\Phi)}{p_1}
\end{equation}
where again $W^\prime(\Phi)=V(\Phi)$. $S^{(0)}$~can be extended to a complete
integral by making the ansatz
\begin{equation}\label{cians}
S(x,\Phi,\lambda,p_1,p_2)=S^{(0)}(\Phi,\lambda,p_1)+
   p_2\left[\Delta(\Phi,\lambda,p_1,p_2)-x\right]\,.
\end{equation}
It is not difficult to show that such a function~$\Delta$ always exists. The
special form of~(\ref{cians}) allows us to evaluate the canonical
transformation~(\ref{ctrafo}) on the constraint surface~$p_2=0$. There we
obtain
\begin{equation}
\astretch{1.5}
\begin{array}{c}
\pi_\Phi=V(\Phi)/p_1\,,\quad \pi_\lambda=p_1 e^\lambda\,,\\
q^1=e^\lambda-W(\Phi)/(p_1)^2\,.
\end{array}
\end{equation}
We cannot compute $q^2$, but this does not matter, as it is purely gauge.

The new coordinates~$(q^1,p_1)$ are gauge independent observables, as one can
easily check that their Poisson brackets with the Hamiltonian vanish.
Furthermore we can relate them with the integration constants~$A,B$ used in
Section~\ref{redode}
\begin{equation}
A=p_1\,,\quad B=(p_1)^2q^1\,.
\end{equation}

\section{Quantization}\label{quant}

Since we have reduced dilaton gravity to the zero energy sector of a
finite-dimensional dynamical system, we can quantize it in a simple way
obtaining a standard Wheeler-DeWitt equation instead of a functional equation.
We choose the usual representation of the momenta in terms of partial
derivatives. The vanishing of the classical Hamiltonian~(\ref{hamg}) yields the
following hyperbolic equation for the wave function~$\bf\Psi(\Phi,\lambda)$
\begin{equation}
\hbar^2\frac{\partial^2{\bf\Psi}}{\partial\Phi\partial\lambda}+
e^\lambda V(\Phi){\bf\Psi}=0\,.
\end{equation}
The simple field redefinition $\mu=e^\lambda$, $\rho=W(\Phi)$ where again
$W^\prime(\Phi)=V(\Phi)$ transforms it into the Klein-Gordon equation
(in characteristic coordinates)
\begin{equation}\label{kg}
\hbar^2\frac{\partial^2{\bf\Psi}}{\partial\rho\partial\mu}+{\bf\Psi}=0\,.
\end{equation}

In order to validate our quantization procedure we compute the semi-classical
limit of this theory using the {\small WKB} approach. Thus we make the
following
ansatz for ${\bf\Psi}$ depending on two real fields~$S,{\cal A}$
\begin{equation}
{\bf\Psi}(\rho,\mu)={\cal A}(\rho,\mu)e^{\frac{i}{\hbar}S(\rho,\mu)}\,.
\end{equation}
(\ref{kg})~yields the following \de
\begin{equation}
\hbar^2{\cal A}_{\rho\mu}+i\hbar({\cal A}S_{\rho\mu}+{\cal A}_\mu S_\rho+
  {\cal A}_\rho S_\mu)-{\cal A}S_\rho S_\mu+{\cal A}=0\,.
\end{equation}
Now we expand both functions in power series in~$\hbar$: ${\cal A}={\cal
A}^{(0)}+\hbar {\cal A}^{(1)}+\dots$ and $S=S^{(0)}+\hbar S^{(1)}+\dots$ In the
classical limit, i.e.\ for $\hbar\rightarrow0$, this leads to
\begin{equation}\label{S0}
S^{(0)}_\rho S^{(0)}_\mu=1\,.
\end{equation}

This is exactly the Hamilton-Jacobi equation~(\ref{hjc}) we obtained in the
last
section (transformed to the new coordinates~$\rho,\mu$) and we can reuse the
incomplete integral~(\ref{inint}). In the new coordinates $\rho,\mu$ the
$\lambda$-$\Phi$ relation~(\ref{lambphi}) derived in Section~\ref{redode} reads
\begin{equation}
A\mu=\frac{\rho+B}{A}\,.
\end{equation}
Identifying $p_1$ with $A$ one can easily that these classical trajectories are
orthogonal with respect to the Minkowski metric to the curves described by
$S^{(0)}=const$.  Thus we obtain the correct classical limit.

{\openup1\jot
For the next terms in the {\small WKB} approximation we obtain the following
\de
s
\begin{eqnarray}
&S^{(0)}_\mu S^{(1)}_\rho + S^{(0)}_\rho S^{(1)}_\mu =0\,,&\\
&S^{(0)}_{\rho\mu} {\cal A}^{(0)}+S^{(0)}_\rho {\cal A}^{(0)}_\mu +
   S^{(0)}_\mu {\cal A}^{(0)}_\rho=0\,,&\\
&S^{(0)}_{\rho\mu} {\cal A}^{(1)}+S^{(1)}_{\rho\mu} {\cal A}^{(0)}+
   S^{(0)}_\rho {\cal A}^{(1)}_\mu + S^{(1)}_\rho {\cal A}^{(0)}_\mu +
   S^{(0)}_\mu {\cal A}^{(1)}_\rho + S^{(1)}_\mu {\cal A}^{(0)}_\rho =0\,.&
\end{eqnarray}
They can be solved easily by introducing the new variables
$2\sigma^\pm=A\rho\pm\mu/A$
\begin{eqnarray}
S^{(0)}(\sigma^+,\sigma^-)&=&\sigma^+ + C\,,\\
S^{(1)}(\sigma^+,\sigma^-)&=&F(\sigma^-)\,,\\
{\cal A}^{(0)}(\sigma^+,\sigma^-)&=&G(\sigma^-)\,,\\
{\cal A}^{(1)}(\sigma^+,\sigma^-)&=&H(\sigma^-)+
  \Bigl[G(\sigma^-)F^{\prime\prime}(\sigma^-)+
        G^\prime(\sigma^-)F^\prime(\sigma^-)\Bigr]\sigma^+
\end{eqnarray}
with an arbitrary constant~$C$ and three arbitrary functions~$F,G,H$.}

\section{Minimally Coupled Matter}\label{matter}

We now couple minimally a matter field~$\psi$ by adding
\begin{equation}
{\cal L}_M=\kappa\sqrt{-g}(\nabla\psi)^2
\end{equation}
with a coupling constant~$\kappa$ to the action~(\ref{dgact}). Its
energy-momentum tensor is given in the conformal gauge~(\ref{conmet}) by
\begin{equation}
\astretch{2}
\begin{array}{l}
\ds T^{00}=T^{11}=\frac{\kappa}{2}e^{-2\lambda}(\psi_t^2+\psi_x^2)\,,\\
\ds T^{01}=-\kappa e^{-2\lambda}\psi_t\psi_x\,.
\end{array}
\end{equation}

Adding again the gauge fixing condition $\lambda_t=0$ it is easy to show that
we
obtain exactly in the same way as before that $\Phi_t=0$ and additionally that
$\psi_t=0$. Thus we can still reduce the field equations to \ode s. Note that
this stems from the fact that there is no coupling between the dilaton field
and
the matter.

The reduced field equations have now the form
\begin{equation}
\astretch{1.3}
\begin{array}{l}
\Phi_{xx}-e^\lambda V(\Phi)=0\,,\\
\lambda_{xx}-e^\lambda V^\prime(\Phi)=0\,,\\
\psi_{xx}=0\,,\\
\Phi_x\lambda_x-e^\lambda V(\Phi)+\kappa\psi_x^2=0\,.
\end{array}
\end{equation}
Again we can identify the last equation with a zero energy condition for the
unconstrained system defined by the Lagrangian
\begin{equation}
{\cal L}_g[\Phi,\lambda,\psi]=
   \Phi_x\lambda_x+\kappa\psi_x^2+e^\lambda V(\Phi)\,.
\end{equation}

Quantizing the Hamiltonian constraint we obtain again a hyperbolic wave
equation
as Wheeler-DeWitt equation
\begin{equation}
\hbar^2\frac{\partial^2{\bf\Psi}}{\partial\Phi\partial\lambda}+
\frac{\hbar^2}{4\kappa}\frac{\partial^2{\bf\Psi}}{\partial\psi^2}+
e^\lambda V(\Phi){\bf\Psi}=0\,.
\end{equation}
In the absence of matter $\Phi$ and $\lambda$ entered the equation on equal
footing. There was no way to decide whether $\Phi+\lambda$ or $\Phi-\lambda$
should be a timelike coordinate in the superspace. Now the sign of~$\kappa$
induces a (2+1)-split of the superspace. However, in general it is not clear
which part of the split is timelike and which spacelike.

\section{Conclusion}

Let us note the different ways to obtain the reduction to ordinary differential
equations in the article by Banks and O'Loughlin~\cite{bl:dilaton} and in our
work. They use right from the beginning the field equations in order construct
the Killing vector~(\ref{kilv}) and find coordinates such that it is rectified.
Thus this approach is specific to dilaton gravity.

We start with a purely {\em mathematical\/} argument. We show that every
two-dimensional metric can be written (locally) in conformal form where the
conformal factor depends only on one of the coordinates. This fact is
independent of any physical model. Particularities of dilaton gravity become
important only when we insert this result into the field equation and it turns
out that in such a coordinate system the dilaton field depends only on this
distinguished coordinate, too.

We also would like to comment on differences in the obtained quantum theories.
They do not consider whether their quantum theory yields the correct classical
limit. Actually, it is easy to see that they would not obtain their classical
model. The latter one depends on three fields, whereas their quantum theory
knows only two degrees of freedom. The field~$g$ used in their parameterization
of the metric simply disappears.

There exists an alternative way to endow the gauge reduced equations of motion
with a Hamiltonian structure. In Section~\ref{hjt} we started with the
second-order system~(\ref{eqmg}). Alternatively one can use the first-order
formulation obtained in Section~\ref{redode} after one integration
\begin{equation}
\astretch{1.3}
\begin{array}{l}
\Phi_x=Ae^\lambda\,,\\
A\lambda_x=V(\Phi)\,.
\end{array}
\end{equation}
These are the Euler-Lagrange equations for the first-order Lagrangian
\begin{equation}\label{lag1}
{\cal L}_1[\Phi,\lambda]=\Phi\lambda_x+Ae^\lambda-\frac{W(\Phi)}{A}\,.
\end{equation}
It is well-known that such a Lagrangian leads directly to generalized Poisson
brackets~\cite{sm:dyn} which can also be considered as Dirac
brackets~\cite{gov:ham}. Applying this formalism to~${\cal L}_1$ yields
\begin{equation}
\{\lambda,\Phi\}=1\,.
\end{equation}
In this description we can thus interpret the dilaton and the conformal factor
as canonically conjugate coordinates! However, we believe that (\ref{lag1})
represents a dubious starting point for a quantization, as $A$ is treated as a
parameter. However, we saw in Section~\ref{hjt} that it can be identified with
a
dynamical variable!

Since we have not been able to find a complete integral of the Hamilton-Jacobi
equation~(\ref{hje}) we could not pursue this argument until the end. We have
not constructed the full canonical transformation which leads to the decoupling
of the Hamiltonian. Otherwise we could have used its regular, gauge-independent
part for the quantization and thus quantize the fully reduced phase space.

Instead we have used a finite-dimensional classical system and imposed from the
outside a gauge symmetry by considering only its zero energy sector. This
symmetry corresponds to the residual gauge freedom left after fixing the gauge
with the condition~$\lambda_t=0$. Then we proceed in the usual way following
Dirac~\cite{dir:qm} by requiring that the wave function is annihilated by an
operator version of the (first-class) constraint.

It appears natural to ask for the relationship between the quantum theory
obtained this way and the one obtained by following the above mentioned
Hamilton-Jacobi procedure. One can expect that they are not equivalent. This
situation is very similar to the quantization of the free relativistic
particle.
The approach we took here corresponds to the covariant quantization. No gauge
fixing is performed and we get a covariant wave function (the Klein-Gordon
equation~(\ref{kg}) is obtained in characteristic coordinates!).

One should probably study in more detail the relation between the residual
gauge
symmetry in the reduced field equations~(\ref{eqmg}) and the symmetry generated
by the constraint~${\cal H}_g=0$. As mentioned in Section~\ref{redode} the
integration constant~$A$ can be changed by a rescaling of~$x$. In the context
of
the field equations we consider this as a gauge transformation. For the system
described by the particle Lagrangian~${\cal L}_g$ this corresponds to a
reparametrization of the evolution parameter~$x$ and is not contained in the
gauge transformations generated by~${\cal H}_g$. Under these transformations
$A=p_1$ remains invariant.

This connection can be made more transparent by using a reparametrization
invariant action. To this end one introduces a new evolution parameter~$\gamma$
and sets $x=X(\gamma)$. This leads to the action (the dot denotes derivatives
with respect to $\gamma$)
\begin{equation}
{\cal L}_g^\prime[\lambda,\Phi,X]=\frac{\dot\Phi\dot\lambda}{\dot X}+\dot X
e^\lambda V(\Phi)\,.
\end{equation}
The original equations of motion are recovered, if one imposes the gauge fixing
condition $X-\gamma=0$. Since this condition depends explicitly on the
evolution
parameter, the gauge fixed Hamiltonian acquires a correction
term~\cite{ev:etd,et:tdgf}. Once this is taken into account, one obtains
exactly
the same quantum theory as we did in Section~\ref{quant}.

Finally, we would like to stress again that applying methods from the formal
theory of \pde s allows us to compute the dimension of the fully reduced phase
space without constructing it. This indicates that these techniques should also
be useful for more complicated models where this construction cannot be
performed explicitly.

This holds especially for systems where one can show that for a full gauge
reduction one must pose in addition initial and/or boundary conditions. For
instance in the case of standard four-dimensional general relativity it is easy
to see that the gauge corrected \cc s cannot be obtained from any system of \de
s, as they do not satisfy all properties of \cc s. This implies that it is not
possible to fix the gauge completely by imposing gauge conditions in the form
of
differential (or algebraic) equations. Nevertheless, one can determine the
arbitrariness of the fully reduced phase space~\cite{wms:diss,wms:sym}.

\section*{Acknowledgments}

{\small WMS}~is grateful for support by a grant of the School of Physics and
Materials, Lancaster University and {\small RWT} for support by a grant in the
{\small EC} Human Capital and Mobility program.

\end{document}